\documentclass[aps,prb,twocolumn]{revtex4}

\usepackage{hyperref}
\usepackage{graphicx}
\usepackage{amsfonts}
\usepackage{amssymb}
\usepackage{amsmath}

\usepackage{color}

\newcommand{\dg}{\dagger}

\newcommand{\bg}{b^{\dg}}
\newcommand{\la}{\langle}
\newcommand{\ra}{\rangle}
\newcommand{\ua}{\uparrow}
\newcommand{\da}{\downarrow}

\font\elevenmib=cmmib10 scaled 1095

\skewchar\elevenmib='177
\newfam\mibfam

\newcommand{\CC}{\mathbb{C}}

\newcommand{\beq}{\begin{equation}}
\newcommand{\eeq}{\end{equation}}

\newcommand{\bra}[1]{\big<#1|}
\newcommand{\ket}[1]{|#1\big>}
\newcommand{\braket}[2]{\big<#1|#2\big>}

\newcommand{\barr}{\begin{eqnarray}}
\newcommand{\earr}{\end{eqnarray}}

\def\nd{^{\vphantom{\dagger}}}
\def\yd{^\dagger}
\mathchardef\sigma="711B

\def\nhat{{\hat n}}

\def\ket#1{{|#1\rangle}}
\def\bra#1{{\langle #1 |}}
\def\braket#1{{\langle #1|#1\rangle}}

\def\seq#1{\langle #1 \rangle}

\def\Paren#1{\left( #1 \right)}		

\begin{document}

\title{AKLT Models with Quantum Spin Glass Ground States} 

\author{C. R.  Laumann}

\affiliation{Department of Physics, Joseph Henry
Laboratories, Princeton University, Princeton NJ 08544, USA}

\author{S. A.  Parameswaran}

\affiliation{Department of Physics, Joseph Henry
Laboratories, Princeton University, Princeton NJ 08544, USA}

\author{S. L. Sondhi}

\affiliation{Department of Physics, Joseph Henry
Laboratories, Princeton University, Princeton NJ 08544, USA}

\author{F. Zamponi}

\affiliation{Laboratoire de Physique Th\'eorique, Ecole Normale Sup\'erieure, 
UMR CNRS 8549, 24 Rue Lhomond, 75231 Paris Cedex 05, France}

\date{\today}

\begin{abstract}

We study AKLT models on locally tree-like lattices of fixed connectivity and
find that they exhibit a variety of ground states depending upon the spin,
coordination and global (graph) topology. We find a) quantum paramagnetic or
valence bond solid ground states, b) critical and ordered N\'{e}el states on
bipartite infinite Cayley trees and c) critical and ordered quantum vector
spin glass states on random graphs of fixed connectivity. We argue, in
consonance with a previous analysis\cite{Laumann:2009p8702}, that {\it all}
phases are characterized by gaps to local excitations. The spin glass states
we report arise from random long ranged loops which frustrate
N\'eel ordering despite the lack of randomness in the coupling strengths.

\end{abstract}

\maketitle


\section{Introduction} 
\label{sec:Intro}

The study of quantum antiferromagnets has proven among the most enduring
themes in modern condensed matter physics. The interplay between frustration
and quantum fluctuations leads such systems to exhibit a variety of
interesting ground states. In this context the lattice models constructed by
Affleck \emph{et. al} \cite{affleck1988vbg,PhysRevLett.59.799} are
particularly useful for they build in a great deal of both these effects using
simple local projectors, which allow their ground states to be determined
analytically. These AKLT models have spins given by $S = \frac{z}{2}M$, where
$M$ is a positive integer, and $z$ the lattice coordination number. In
principle, they may be defined on any graph, but in practice one usually
maintains fixed connectivity in order to have the same spin on each site. The
associated ground states have the added feature that their wavefunctions can
be written in Jastrow (pair product) form, which allows us to view their
ground-state probability densities as Boltzmann weights corresponding to a
nearest neighbor Hamiltonian for classical vector spins on the \emph{same}
lattice. Using this unusual quantum-classical equivalence one can understand
many properties of the states by studying the associated classical model.

The initial construction of the AKLT models was motivated by the search for
quantum disordered states in low dimensions. This works only too well: in
$d=1$ and $d=2$ the mapping to finite temperature classical models discussed
above ensures, by the Mermin-Wagner theorem, that {\it all} cases lead to
quantum paramagnetic or valence bond solid ground states. In $d>2$ this is no
longer true and a computation is needed to decide which models order and which
do not. In a recent paper, Parameswaran, Sondhi and Arovas
\cite{Parameswaran:2009p8701} showed via Monte Carlo simulations and
mean-field arguments that AKLT models on the diamond and pyrochlore lattices
exhibit quantum-disordered ground states for small spin sizes while on the
cubic lattice all models exhibit N\'{e}el order.

In this paper we take this exploration of higher dimensional AKLT models in a
different direction---we study them on locally tree-like lattices of fixed
connectivity $z$, which are known to physicists as Bethe lattices. Here we
shall consider two physically distinct systems. The first, is the Bethe
lattice constructed as the limit of a family of Cayley trees. This
construction yields a system with a finite surface to volume ratio and without
loops. The second is a typical member of the ensemble of random graphs of
fixed connectivity. These graphs are locally tree-like in the thermodynamic
limit; however they also have long loops of logarithmically divergent size.
These loops of both even and odd lengths introduce topological frustration
into the system. The two constructions of locally tree-like lattices yield
different physics.

For the infinite Cayley tree, we exhibit an exact solution using the
quantum-classical correspondence. Specifically, we use a generalized transfer
matrix technique to obtain exact solutions for various statistical quantities
in the ground state of the tree. We note that the AKLT model
on the Bethe lattice has been studied before directly within the quantum
formalism \cite{Fannes:1992p5750}; we suspect that readers will find
our solution simpler. We find one quantum disordered state ($M=1$ on the $z=3$
tree) and two that are critical ($M=2$ on $z=3$ and $M=1$ on $z=4$), in that
the correlation functions decay exponentially at precisely the rate required
to balance the exponential growth of the graph. All other cases exhibit N\'eel
order. We address the question of whether the bulk excitations are gapless in
cases when the AKLT wavefunction has critical or N\'eel correlations. We find,
perhaps surprisingly, that the system is always gapped to \emph{local}
excitations, and that the only gapless excitation is a global one connecting
the different broken-symmetry ground states. We connect this to related work
in Ref.~\onlinecite{Laumann:2009p8702}, in which we conjecture that this is a
generic feature of symmetry breaking quantum models on the Bethe lattice,
related to the spectrum of the graph Laplacian.

On random graphs of fixed connectivity, N\'eel ordering in the companion classical model is frustrated by the presence of the long loops. To study this case, we appeal to the cavity techniques familiar from the theory of classical disordered systems. These have been applied recently to a variety of discrete statistical mechanical problems on random graphs and there is much evidence that the (approximate) techniques are on solid ground. From this analysis, we conclude that for $z \le 10$ there are disordered states at small spin and spin glass ground states at larger spin as well as a couple of cases where the state is critical. For $z > 10$ all AKLT models have ground states with spin glass order. By spin glass order, we mean states with fixed but randomly oriented local magnetizations and that the set of such states is larger than those connected by global rotations alone. We argue that the spectrum of local excitations above the pure states in this set is again gapped.

Of our various results we would especially like to flag these last mentioned.
The nature of quantum glass phases is a subject of much interest -- especially
as to how much of the elaborate framework of the classical subject may be
lifted into the quantum world. The AKLT construction provides a direct line of
approach to this problem and does so using Hamiltonians without random
couplings but from graph disorder alone.

This paper is organized as follows: in Section \ref{sec:AKLTIntro}, we introduce the AKLT model on an arbitrary graph via the Schwinger boson formalism. We proceed to construct a companion classical model that captures the structure of the ground state wavefunction by introducing a basis of $SU(2)$ coherent states. In Section \ref{sec:cont-spin-model}, we develop transfer-matrix technology to solve the companion classical model on the (bipartite) Bethe lattice exactly, and obtain the transition temperature and correlation functions in the paramagnetic and N\'eel-ordered phases. In Section \ref{sec:nogap} we investigate the energy gap using a variational \emph{ansatz} for the excited states. Finally, in Section \ref{sec:reg-rand-graph} we consider the extension of this analysis to the spin glass transition expected on regular random graphs and consider some of the quantum consequences of the classical glassy phase.


\section{AKLT States: A Brief Review} 
\label{sec:AKLTIntro}

The central idea of the AKLT approach \cite{PhysRevLett.59.799} is to
use quantum singlets to construct correlated quantum-disordered
wavefunctions, which are eigenstates of local projection
operators. One can then produce many-body Hamiltonians using
projectors that extinguish the state, thereby rendering the parent
wavefunction an exact ground state, often with a gap to low-lying
excitations. A general member of the family of valence bond solid
(AKLT) states can be written compactly in terms of Schwinger bosons
\cite{PhysRevLett.60.531}:
\begin{equation}
\label{eq:AKLT-def}
|\Psi(M)\ra = \prod_{\la ij\ra} \left(\bg_{i\ua}\bg_{j\da} - \bg_{i\da}\bg_{j\ua}\right)^{\!M} |\,0\,\ra\ .
\end{equation}
This assigns $M$ singlet creation operators to each link $\la ij \ra$ of an underlying lattice. The total boson occupancy per site is
given by $z M$, where $z$ is the lattice coordination number, and the
resultant spin on each site is given by $S = \frac12 z M$. Given any regular graph, the above construction defines a family of AKLT states labeled by the size of their spins $S = \frac12 z M$. For more details regarding this construction and the corresponding Hamiltonians, see Ref.~\onlinecite{Parameswaran:2009p8701}.

The AKLT states have a convenient representation in terms of ${SU}(2)$
coherent states, as first shown in Ref.~\onlinecite{PhysRevLett.60.531}.  In terms of the Schwinger
bosons, the normalized spin-$S$ coherent state is given by $|
\nhat\ra=(p\,!)^{-1/2}\,(z\nd_\mu b\yd_\mu)^p\,|\,0\,\ra$, where
$p=2S$, with $z=(u\, , \, v)$ a $\mathbb{CP}^1$ spinor, with
$u=\cos(\theta/2)$ and $v=\sin(\theta/2)\,e^{i\varphi}$.  The unit
vector $\nhat$ is given by $n^a=z\yd \sigma^a z$, where ${\vec\sigma}$
are the Pauli matrices.  In the coherent state representation, the
general AKLT state wavefunction is the pair product $\Psi(\{\hat n_i \})=
\langle\{\hat n_i \}|\Psi\rangle=\prod_{\la
  ij \ra} (u\nd_i\,v\nd_j - v\nd_i\,u\nd_j)^M$.  Following
Ref.~\onlinecite{PhysRevLett.60.531}, we may write
$|\Psi(\{\hat n_i \})|^2\equiv\exp(-\beta H_{\rm cl})$ 
as the Boltzmann weight for a
classical $\textsf{O}(3)$ model with Hamiltonian
\begin{equation} \label{eq:classicalAKLT}
H\nd_{\rm cl}=-\sum_{\la ij \ra} \ln\!\bigg({1-\nhat_i\cdot\nhat_j\over 2}\bigg)\ ,
\end{equation}
at inverse temperature $\beta=M$.  All equal time quantum correlations in the
state $|\Psi\ra$ may then be expressed as classical, finite
temperature correlations of the Hamiltonian $H\nd_{\rm cl}$.

Some immediate consequences of this quantum-to-classical equivalence were
noted in Ref.~\onlinecite{PhysRevLett.60.531}. On one and two-dimensional
lattices, the Hohenberg-Mermin-Wagner theorem precludes long-ranged order at
any finite value of the discrete quantum parameter $M$. In three dimensions,
there is no \emph{a priori} reason to rule out long-range order. In fact, as
shown in Ref.~\onlinecite{Parameswaran:2009p8701}, the simple cubic lattice
has no quantum-disordered states at any $M$, while the diamond lattice has a
single such state for $M=1$; on frustrated lattices such as the pyrochlore,
such states are believed to exist for many values of $M$.

As is evident from \eqref{eq:AKLT-def}, we may define the AKLT states on an arbitrary graph; if the graph has fixed connectivity $z$, then the resulting model has the same spin on each site. On graphs with a boundary, this is not automatic, since the boundary sites will have fewer neighbors $z'$. There are several ways to deal with this boundary effect. The first is to work with a system with a lower spin on the boundary: in that case $S' = \frac12 z'M$. The quantum state of this non-homogeneous system is unique. Another option is to add $(z-z')$ additional Schwinger bosons of either flavor to the edge sites; there is not a unique way in which to do this, leading to a multitude of degenerate ground states classified by the state of the boundary spins. When translated to the companion classical model, the latter option can be viewed as connecting each of the boundary spins to $(z-z')$ fixed spins, each with an orientation specified by the behavior of an independent spin-$1/2$ degree of freedom; thus, the different degenerate states of the homogeneous AKLT model on a graph with boundary can be understood by choosing different fixed boundary conditions for spins in an additional, outer ring of leaf spins. Finally, one can opt to get rid of the boundary by, for instance, taking periodic boundary conditions on a Euclidean lattice. As we will discuss in Sec.~\ref{sec:reg-rand-graph}, when generalized to tree-like graphs, this approach leads to a spin glass phase in the companion model and thus provides a non-trivial new quantum spin glass to the AKLT phase diagram.


\section{Transfer Matrix Solution of the Classical Problem on Trees} 
\label{sec:cont-spin-model}

The classical Hamiltonian that describes the properties of the
ground-state wavefunction of an AKLT model with singlet index $M$,
written in the basis of $SU(2)$ coherent states is given by
\begin{equation}
  \label{eq:aklt-ham}
  \beta H_\text{cl} = - M \sum_{\seq{i,j}}\log \left[1-\nhat_i \cdot \nhat_j\over 2\right]
\end{equation}

Before we proceed, we note that on bipartite graphs we can perform a gauge transformation
by flipping every spin at odd depth to obtain a ferromagnetic
model. This gives us
\begin{equation}
  \label{eq:aklt-ham2}
  \beta H_\text{cl} = - M \sum_{\seq{i,j}}\log \left[1+\nhat_i \cdot \nhat_j\over 2\right]
\end{equation}
As usual in the treatment of tree models, we consider the statistical state
$\psi^0(\nhat_0)$ (marginal distribution) of a cavity spin $\nhat_0$ at the
root of a branch of the tree. This unnormalized distribution can be
found in terms of the cavity states of its $z-1$ neighbors by summation:
\begin{widetext}
\begin{eqnarray}
  \label{eq:cont-cavity-iter}
  \psi^0(\nhat_0) & = & \int D\nhat_1\cdots D\nhat_{z-1}\,
  T(\nhat_0,\nhat_1)\psi^1(\nhat_1) \cdots
  T(\nhat_0,\nhat_{z-1}) \psi^{z-1}(\nhat_{z-1}) \nonumber
  \\&=& \int D\nhat'_1 \cdots \nhat'_{z-1} M(\nhat_0; \nhat'_1,\cdots \nhat_{z-1}') \int D\nhat_1\cdots D\nhat_{z-1}\,
  T(\nhat'_1,\nhat_1)\psi^1(\nhat_1) \cdots
  T(\nhat'_{z-1},\nhat_{z-1}) \psi^{z-1}(\nhat_{z-1})
\end{eqnarray}
\end{widetext}
where
\begin{equation}
  \label{eq:trans-matrix-def}
  T(\nhat_0,\nhat_1) = e^{\beta \log \left[1+\nhat_0 \cdot \nhat_1\over 2\right]}=\left(1+\nhat_0 \cdot \nhat_1\over 2\right)^\beta
\end{equation}
is the transfer matrix of the AKLT model and
\begin{equation}
  \label{eq:cont-merge-matrix}
  M(\nhat_0; \nhat_1',\cdots \nhat_{z-1}') = \delta(\nhat_0 -\nhat_1')\cdots\delta(\nhat_0-\nhat_{z-1}')
\end{equation}
is the merge matrix.

The merge matrix $M$
defines a multilinear map from the $z-1$ state spaces of the neighbor spins to the state space of the root. This
lifts naturally to the appropriate complexified tensor product spaces %
\footnote{A few comments on the nature of the classical statistical
  state space of a vector spin are in order, as we have so cavalierly
  complexified and tensored it into a much more quantum mechanical
  looking system. Physical cavity distributions $\psi(\nhat)$ must be
  real, normalizable, nonnegative functions on the sphere. By its construction as a marginalization (summing out)
  procedure, equation \eqref{eq:cont-cavity-dirac} must produce such a
  physical output given physical inputs, even though we have extended
  it over $\CC$.  A more important subtlety
  arises in the normalization of states -- the standard $L^2$ norm
  associated with the Dirac inner product is not necessarily $1$ for a
  properly normalized probability distribution. Since the
  probabilistic $L^1$ norm is incompatible with the Hilbert space
  structure, it is much simpler to work with unnormalized vectors and
  keep in mind that a probabilistic interpretation only applies in the
  standard basis.} %
and thus we will find it natural to write the merge and transfer
operations abstractly using Dirac notation:
\begin{widetext}
\begin{eqnarray}
  \label{eq:cont-merge-dirac}
  M &= & \int D\nhat_0 D\nhat_1'\cdots D\nhat_{z-1}'\,\delta(\nhat_0 -\nhat_1')\cdots\delta(\nhat_0-\nhat_{z-1}')
  \ket{\nhat_0}\bra{\nhat_1'}\cdots\bra{\nhat_{z-1}'} \nonumber \\
  & = & \int D{\nhat}\, \ket{\nhat}\bra{\nhat}\cdots\bra{\nhat}
\end{eqnarray}
\end{widetext}
and,
\begin{equation}
  \label{eq:cont-transfer-dirac}
  T = \int D\nhat D\nhat'\, T(\nhat,\nhat')\,\ket{\nhat}\bra{\nhat'}.
\end{equation}
Thus, equation \eqref{eq:cont-cavity-iter} becomes
\begin{equation}
  \label{eq:cont-cavity-dirac}
  \ket{\psi^0} = M \Paren{T\ket{\psi^1} \otimes \cdots \otimes T\ket{\psi^{z-1}}}.
\end{equation}

We now focus on the stability of the paramagnetic state against
N\'eel ordering. Hence, we have to use boundary conditions that are consistent
with this kind of ordering. As discussed at the end of Section~\ref{sec:AKLTIntro}, 
one can either use free boundary spins with lower $S$, or connect the boundary spins to some
fixed additional spins: in the latter case, the additional spins must all have the same orientation
to allow the N\'eel ordering. This remark is particularly important because, as we will see later in
Section~\ref{sec:reg-rand-graph}, on a random regular graph the boundary conditions on any
given tree-like subregion are fixed self-consistently, and in general are not consistent with N\'eel
ordering, leading to a disordered spin glass state.
Assuming uniform boundary conditions, we obtain the unique state at
depth $d-1$ by merging the $z-1$ states at level $d$ using the $T$ and
$M$ operators:
\begin{equation}
  \label{eq:cont-cavity-cayley}
  \ket{d-1} = M (T\ket{d})^{\otimes (z-1)}.
\end{equation}

The natural basis to work in is that of states with definite angular
momentum, i.e. states $\ket{l\,m}$, which are eigenstates of the
angular momentum operators $L^2, L_z$.
In the coordinate basis, these are simply the
spherical harmonics, and as shown in Appendix  \ref{appsec:Tmatrix}, they are eigenstates of the transfer matrix with eigenvalue $\lambda_l$.

It remains for us to understand exactly how the merge operation acts
in the angular momentum basis. If we insert resolutions of the
identity in the angular momentum basis into
\eqref{eq:cont-merge-dirac}, we obtain
\begin{widetext}
	\begin{eqnarray}
	\label{eq:cont-merge-angmtm}
	M &=& \int D{\nhat}\, \ket{\nhat}\bra{\nhat}\cdots\bra{\nhat}\nonumber
	\\&=& \int D{\nhat}\sum_{l_0,m_0}\sum_{l_1,m_1}\cdots\sum_{l_{z-1}, m_{z-1}} \ket{l_0,m_0}\langle l_0,m_0|\nhat\rangle \left(\langle \nhat |{l_1,m_1}\rangle \bra{l_1,m_1}\otimes\cdots\otimes\langle\nhat|{l_{z-1},m_{z-1}}\rangle\bra{l_{z-1},m_{z-1}}\right)\nonumber
	\\&=& \sum_{l_0,m_0}\sum_{l_1,m_1}\cdots\sum_{l_{z-1}, m_{z-1}} \ket{l_0,m_0}\left(\bra{l_1,m_1}\otimes \cdots \otimes \bra{l_{z-1},m_{z-1}}\right) \int D{\nhat}\, Y_{l_0}^{m_0*}(\nhat) Y_{l_1}^{m_1}(\nhat)\cdots Y_{l_{z-1}}^{m_{z-1}}(\nhat)
\end{eqnarray}
\end{widetext}

For the case $z=3$, the integral in \eqref{eq:cont-merge-angmtm} is
simply the Clebsch-Gordan coefficient that characterizes the fusion of
two $SU(2)$ spins. For higher values of
$z$, this is the appropriate generalization of the Clebsch-Gordan
coefficient describing the fusion of $(z-1)$ $SU(2)$ spins. Thus we see that the merge operation, when
written in the angular momentum basis, has a natural interpretation as
the fusion rules for the $O(3)$ symmetry group.

The paramagnetic state - here represented in Fourier space by the
$\ket{00}$ state - is always a fixed point: it is an eigenstate of the
$T$-matrix, and the fusion of any number of $\ket{00}$ states is again
a $\ket{00}$ state. We proceed via linear stability analysis: we
introduce a perturbation into a state that is not uniformly weighted
on the sphere, and see if this grows or shrinks under the iteration
procedure. We note that we can decompose any such state into spherical
harmonics, and so we write $\ket{d} = \ket{00} +\epsilon\sum_{l\neq0,
  m}^{\infty} c_{lm}\ket{l\,m}$ into \eqref{eq:cont-cavity-cayley} to
obtain
\begin{eqnarray}
	\label{eq:cont-merge-floweq}
	\ket{d-1} &=& M\left(\lambda_0\ket{00} + \epsilon \sum_{l\neq 0, m} \lambda_l c_{lm}\ket{l\,m}\right)^{\otimes (z-1)}	     \nonumber\\
	&=& \lambda_0^{z-1} \ket{00} \nonumber\\&\left.\right.& +\epsilon (z-1) \lambda_0^{q-2} \sum_{l\neq 0, m}\lambda_l c_{lm}\ket{l\,m} + \mathcal{O}(\epsilon^2)
\end{eqnarray}
where we have used the fact that fusing any number of $\ket{00}$ states
with an $\ket{l\, m}$ state results in an $\ket{l\,m}$ state. We
renormalize to leading order and find that the iterated state is, to
linear order
\begin{equation}
	\label{eq:cont-iter-state}
	\ket{d-1} =  \ket{00} + \epsilon (z-1) \sum_{l\neq 0, m} \frac{\lambda_l}{\lambda_0}c_{lm}\ket{l\,m} + \mathcal{O}(\epsilon^2)
\end{equation}

The perturbation is irrelevant (shrinks under iteration) if the
coefficient of the linear term is less than $1$, and relevant if it is
greater than 1.  The critical point is reached when
\begin{equation}
	\label{eq:cont-crit-pt}
	\frac{\lambda_l}{\lambda_0} = \frac{1}{z-1}
\end{equation}
for any $l$. Using the temperature dependent expression
(\ref{eq:finaleval}) for the $\lambda_l$, one can show that the dipole
instability ($l=1$) is the first one encountered as the temperature is
lowered, and therefore sets the transition temperature.

Using the results of Appendix \ref{appsec:Tmatrix} (replacing $\beta$
by the singlet parameter $M$), we obtain
\begin{equation}
	\label{eq:AKLTTc}
	M_c = \frac{2}{z-2}
\end{equation}
We see that for $z=2, 3, 4$, $M_c = \infty, 2, 1$, while for all other
values, $M_c < 1$. Since $M$ must be a positive integer, we see that
for the chain ($z=2$) all values of $M$ correspond to
quantum-disordered states (which follows from the Mermin-Wagner
theorem and the original AKLT result\cite{affleck1988vbg}) whereas for $z=3$, the $M=1$ state is disordered while the
$M=2$ state is critical, and finally for $z=4$, the $M=1$ state is
critical. Bethe lattices of higher connectivity will always have
ordered AKLT ground states for any value of $M$. See Fig.~\ref{fig:figs_fig-akltpd}.

Finally, we consider the correlation function $\langle \hat n_0 \cdot \hat n_d \rangle$ within the paramagnetic phase. This is given by considering the response of $\langle \hat n_0 \rangle$ to a field on $\hat n_d$ -- which is  the same as asking how the dipole $l=1$ perturbation propagates along a chain of length $d$ in a background of trivial $l=0$ cavity states. This immediately implies
\begin{equation}
	\label{eq:corrfunc}
	\langle \hat n_0 \cdot \hat n_d \rangle \propto \left(\frac{\lambda_1}{\lambda_0}\right)^{d} = \left( \frac{M}{M+2} \right)^d
\end{equation}
Notice that this implies that the naive correlation length never diverges -- as usual with tree models, phase transitions occur when the correlation decays at the same rate as the growth of the lattice. For a slightly more detailed calculation, see Appendix~\ref{sec:stabSG}.

As an aside, we note that we can use the same generalized transfer
matrix technique to obtain the transition temperature for the
Heisenberg model, a result first obtained by Fisher
\cite{Fisher:1964p5715} using a different method. This serves as a
test of the technique proposed here.


\section{Variational Bounds on the Gap} 
\label{sec:nogap}

We now perform a variational computation of the gap to excitations in the critical model, similar to the Single-Mode Approximation (SMA) discussed in Ref.~\onlinecite{PhysRevLett.60.531}.  The central idea of the SMA is to construct an excitation orthogonal to the ground state by acting on it with a local operator, and then to reduce the energy of this excitation by delocalizing it, thereby decreasing its kinetic energy.  A variational bound on the energy gap is given by
\begin{equation}\
0\leq \Delta \leq \Delta_{SMA} =\frac{\bra{\Psi_{SMA}} H-E_0\ket{\Psi_{SMA}}}{\braket{\Psi_{SMA}}}
\end{equation}

This approach is designed to optimize the energy due to the off-diagonal matrix elements in the excited sector, which will be proportional to the usual graph Laplacian for a nearest neighbor model. On the Bethe lattice, the spectrum of the Laplacian is unusual. As argued in Ref.~\onlinecite{Laumann:2009p8702}, there is necessarily a gap to  hopping excitations on tree-like graphs despite the existence of symmetry related  ground states. Thus, in some sense the SMA calculation is doomed to failure as it will never be able to close this gap. Nonetheless, it is interesting to see how this plays out in an exact model.

We consider a rooted Cayley tree, and in order to restrict ourselves to studying excitations confined to the bulk as the size of the tree grows we suppress excitations far from the center using an infrared regulator $\lambda$. We therefore study variational wavefunctions of the form
\begin{equation}
 \ket{\Psi_\text{SMA}}=\ket{\lambda} = \sum_{i=1}^{N}  u_\lambda (i) S_{i}^z\ket{\Psi}
\end{equation}
where $u_\lambda(i)$ is a function only of the depth $\rho_i$ of site $i$ referenced to the root of the Cayley tree.
The SMA gap is 
\begin{equation}
\Delta_{SMA} =  \lim_{\lambda\rightarrow 0} \left[\lim_{N\rightarrow\infty}\frac{\bra{\lambda} H-E_0\ket{\lambda}}{\braket{\lambda}}\right]
\end{equation}
After some algebra, we may write (with the understanding that we always take $N$ to infinity \emph{before} taking $\lambda$ to zero):
\begin{eqnarray}
 \Delta_{SMA} = \lim_{\substack{N\rightarrow \infty\\ \lambda\rightarrow 0}} \left[\frac{\sum_{i,j} u_\lambda (i) u_\lambda (j)  \langle [S_i^z,[H,S_j^z]]\rangle_\Psi }{\sum_{i,j}   u_\lambda (i) u_\lambda (j)  \langle S_i^z S_j^z \rangle_\Psi}\right]
\end{eqnarray}

Asserting homogeneity of the graph, and noting that the Hamiltonian is a function only of $S_i\cdot S_j$ where $i$ and $j$ are nearest neighbors, we may re-express this as
\begin{eqnarray}\label{gen-sma}
 \Delta_{SMA} =  \lim_{\substack{N\rightarrow \infty\\ \lambda\rightarrow 0}}\left[f\times\frac{\sum_{ i,j} u_\lambda (i)\mathcal{A}_{ij}u_\lambda (j) }{\sum_{i,j}   u_\lambda (i)  \gamma^{|i-j|}u_\lambda (j) }\right]
\end{eqnarray}
where $f = \frac14 \langle [S_i^z-S_j^z,[H,S_i^z-S_j^z]]\rangle_\Psi$, and we denote by $\mathcal{A}$ the adjacency matrix of the Cayley tree; the correlations in the ground state are always exponential, and go as $\gamma^{|i-j|}$, where $\gamma \rightarrow \frac{1}{z-1}$ from below as we approach criticality. We have therefore reduced the problem of the SMA on the Bethe lattice to understanding (i) the spectrum of the graph Laplacian (the adjacency matrix up to a sign) and (ii) the behavior of the ground-state correlations.

Our choice for a variational ansatz is to take $u_{\lambda}(i) = e^{-\lambda\rho_i^2}$. This is motivated by the fact that the number of sites at a given distance from the center grows exponentially, and therefore in order to remain in the bulk of the tree, we need to cut off the wavefunction faster than exponentially.\footnote{In principle, an exponential regulator with a sufficiently fast decay constant also works. The calculation proceeds in a similar fashion but the interpretation is more complicated and no more enlightening.} 
We perform the summations by converting the sum over sites into a sum over depths, approximating the sums by integrals and using steepest-descent. We find that, at criticality, the gap is nonvanishing:
\begin{eqnarray}\label{crit-gen-sma}
 \Delta_{SMA}^G \sim \frac{f}{8} z\left[ \frac{z-2\sqrt{z-1}}{z-1+\frac{4\sqrt{z-1}}{\log{(z-1)}}}\right]
\end{eqnarray} 
Excitations constrained to live in the bulk are therefore always gapped, even at criticality. The factor of $z - 2\sqrt{z-1}$ is precisely the spectral gap for bulk excitations on a Cayley tree\cite{Laumann:2009p8702}. A state where $u_\lambda(i)$ is independent of position must be gapless in the broken-symmetry phase of the model, since it connects the different broken-symmetry ground states. We cannot recover this state through a correctly regulated calculation in the chosen order of limits, however.

There is a straightforward physical argument for this gap. By our choice of variational ansatz, we cut off the excitation at some depth $D \sim 1/\sqrt{\lambda}$. On a Euclidean lattice of dimension $d$, this costs a surface energy $\propto D^{d-1}$ which is normalized by the weight of the wavefunction $\propto D^{d}$. As $D\to\infty$ ($\lambda\to 0$), the SMA gap therefore vanishes as $D^{-1}$. On tree-like lattices, both the surface area and the bulk normalization scale as $(z-1)^D$; the boundary is always a finite fraction of the bulk. Thus, the ratio remains finite as $D\to\infty$ and the gap cannot close \cite{Hoory:2006qd,Laumann:2009p8702}.


\section{AKLT model on regular random graphs: frustration and the spin glass state} 
\label{sec:reg-rand-graph}

We now consider the same model on a {\it regular random graph}~\cite{RandomGraphs} of connectivity $z$. The ensemble of these graphs is constructed by assigning uniform probability to all
possible graphs of $N$ vertices, such that each vertex is connected to {\it exactly} $z$ links. There are several reasons why statistical models defined on
this ensemble of graphs are interesting:
\begin{enumerate}
\item
A central property of this ensemble~\cite{RandomGraphs} is that typical lattices are locally tree-like; their loops have a length diverging logarithmically with the size $N$ of the system;
this implies that one can develop a method to solve statistical models on these graphs based on the same recurrence equations that are exact on trees. This is known as
the {\it cavity method} \cite{MP01}.
\item
Despite being locally tree-like, they do not have any boundary, all sites playing statistically the same role (in the same way as periodic boundary conditions impose
translation invariance on a finite cubic lattice).
Moreover, the free-energy of regular random graph models is self-averaging with respect to their random character in the thermodynamic
limit. In other words for large enough $N$ a single sample is a good representative of the ensemble average.
\item
Typical graphs are characterized by many large loops of even and odd length;
this strongly frustrates the antiferromagnetic ordering, which gives way to a spin glass phase instead~\footnote{See Ref.~\onlinecite{MP01} for a general discussion and Ref.~\onlinecite{KZ08} for the explicit computation of
the phase diagram of a classical Ising antiferromagnet.}.
\end{enumerate}
In the following we will be particularly interested in the implications of the last point for the quantum problem.

The reasoning outlined in section~\ref{sec:AKLTIntro} clearly applies to the random graph model, whose AKLT ground state is therefore described by a classical Hamiltonian
of the form (\ref{eq:aklt-ham}), where the pairs $\langle i,j \rangle$ are connected by a link of the random graph.
The main difference between the tree model and the random graph model is that the recurrence equation (\ref{eq:cont-cavity-iter}) now does not hold for the full graph; it only
holds for a tree-like subregion of the graph, and has to be initialized using the boundary values of the $\psi^i(\hat n_i)$ that are determined by the summation over the
rest of the graph. 
In other words, the recurrence on the subregion is initialized from random self-consistent boundary conditions, determined by the rest of the system: these boundary conditions
are not consistent with N\'eel ordering, which is therefore frustrated, as discussed in Section~\ref{sec:cont-spin-model} above.
However, since the tree-like subregions grow in size when $N\to\infty$, equation (\ref{eq:cont-cavity-iter}) must be iterated a very large number of times.
One can classify the different phases of the system by studying its fixed points~\cite{MP01,MM09}.

\begin{figure}[tbp]
	\centering
		\includegraphics{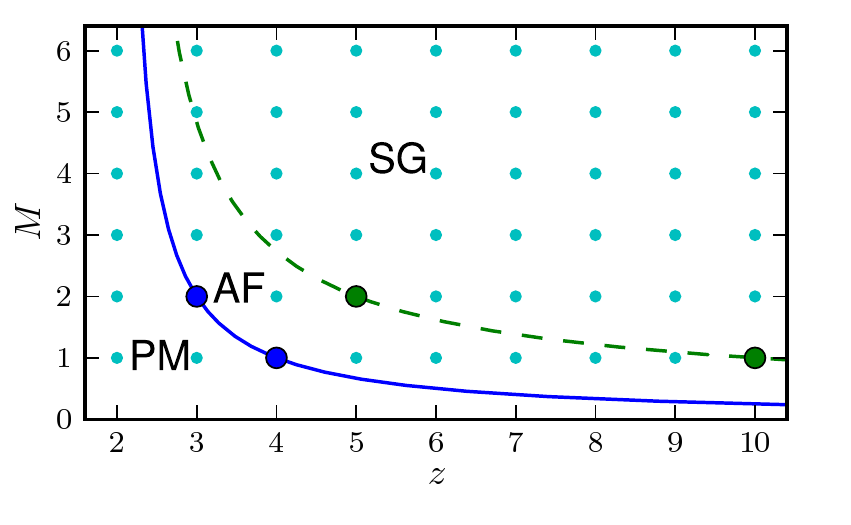}
	\caption{Phase diagrams for AKLT model with singlet parameter $M$ on tree-like lattices with coordination $z$. On the Cayley tree the transition is from paramagnetic (PM) to N\'eel-ordered (AF) phase at the solid blue line with no spin glass. On regular random graphs the transition is from the paramagnetic to spin glass ordered (SG) phase at the dashed green line --- there is no antiferromagnet. The models with Bethe lattice critical correlations are labeled with large dots.}
	\label{fig:figs_fig-akltpd}
\end{figure}

To calculate the stability of the paramagnetic solution against spin glass ordering we observe that the spin glass transition is signaled, as usual, by the divergence of the classical spin glass susceptibility~\cite{BY86}:
\beq
\chi_{SG} = \frac1N \sum_{ij} [ \langle \hat n_i \cdot \hat n_j \rangle ]^2 \ .
\eeq
The details are discussed in Appendix~\ref{sec:stabSG};
the result is that $\chi_{SG}$ is finite if for all $l$
\begin{equation}
	\label{eq:cont-crit-SG}
	\frac{\lambda_l}{\lambda_0} \leq \frac{1}{\sqrt{z-1}} \ .
\end{equation}
Once again the instability originates in the $l=1$ sector and occurs at
\begin{equation}
	\label{eq:AKLTT_SG}
	M_{SG} = \frac{2}{\sqrt{z-1}-1}
\end{equation}
We see that $M_{SG}=\infty$ for $z=2$, $M_{SG}=4.828$ for $z=3$ (hence the system is a spin glass for $M \geq 5$),
$M_{SG}=2$ for $z=5$ and $M_{SG}=1$ for $z=10$ (in these cases the system is critical),
and it is smaller than 1 for any $z>10$. See Fig.~\ref{fig:figs_fig-akltpd}.

With the N\'eel-ordered phase suppressed, the quantum paramagnet extends further in the $z-M$ plane than on the Cayley tree models. Nonetheless, it is clear that the paramagnetic solution develops an instability to spin glass ordering at large $M$. What are the properties of this low temperature phase? Most of the detailed work on classical spin glasses has focussed on discrete models. The AKLT mapping provides a classical \emph{vector} model with weakly divergent interactions whose glass phase has not yet been studied. Nonetheless, most of the qualitative features of the classical multiple-valley picture should still hold and these provide an intriguing scenario for the quantum system. The classical Gibbs measure decomposes into a collection of clustering pure states $\alpha = 1\ldots \mathcal{N}$ with essentially disjoint support. In each of these states, the $\psi_\alpha^i(\hat n_i)$ -- and thus the local magnetizations -- are macroscopically different. This strongly suggests that the quantum AKLT ground state  $| \Psi \rangle$ itself is a superposition over a collection of macroscopically distinct degenerate ground states $|\Psi_\alpha\rangle$ each of which corresponds to one of the classical clustering states. 

While we believe that the above picture holds in general, a rigorous derivation is problematic. In the following we will attempt to justify it in more detail and point out some of the subtleties that must be dealt with. First, there does not yet exist a detailed study of the classical vector model \footnote{Although there is a replica symmetric treatment of a related model with the additional complication of random interactions. See Ref.~\onlinecite{Coolen:2005kx}.} -- for the purposes of this paper, we shall assume that, modulo the $\textsf{O}(3)$ global symmetry, the qualitative behavior is that of the better studied Ising antiferromagnet on a regular random graph~\cite{KZ08}. 

By analogy to this model, two different phases exist: at high temperatures, the stable phase is a paramagnet
where $\psi^i(\hat n_i) = \psi(\hat n) = 1$ is the same for all sites, and it is the unique fixed point of Eq.~(\ref{eq:cont-cavity-iter}).
At low temperatures, the stable phase is a spin glass, characterized by the existence of
many \emph{non-symmetry} related pure states labeled by $\alpha = 1, \ldots, {\cal N}$ within which connected spatial correlators vanish.
This is related to the existence of
many different fixed points of Eq.~(\ref{eq:cont-cavity-iter}), and reflects the decomposition of the thermodynamic Gibbs measure $P = \exp(-\beta H_{cl})/Z$ as follows:
\beq\label{eq:PSD}
P(\{ \hat n \}) = \sum_{\alpha=1}^{\cal N} w_\alpha \int dg\, P_\alpha(\{ g \cdot \hat n \})
\eeq
where the $P_\alpha$ are representative classical pure state measures. By integrating over $g$, the $\textsf{O}(3)$ of global rotations of spin space, we account for the continuous family of symmetry related pure states associated to each representative state $\alpha$. One can access these representative pure states by adding a uniform infinitesimal field to the classical model, but as the quantum AKLT state is a singlet, we prefer to work without explicitly breaking this symmetry~\cite{BY86}.

The Ising spin glass models with two-body interactions which have been studied, such as the Sherrington-Kirkpatrick model~\cite{Beyond} 
and the random graph antiferromagnet~\cite{KZ08}, are characterized by a continuous spin
glass transition\footnote{Other mean field spin glass models, 
such as the $p$-spin model, can have a number of states scaling exponentially in system size. 
These models show a discontinuous spin glass
transition. However, antiferromagnetic models with two-body interactions usually do not show this phenomenology, 
therefore we will not investigate
this transition in this paper.} with a \emph{finite} collection of pure states throughout the spin glass 
phase\footnote{See Ref.~\onlinecite{Beyond}, and in particular the reprint on page 226, for a more detailed discussion of this
delicate statement.}. 
We shall therefore assume that this is true of our collection of representative pure states. Indeed, all we will need is that 
${\cal N}$ grows at most polynomially in $N$ as the thermodynamic limit is taken.

The pure state decomposition (\ref{eq:PSD}) has striking consequences for the structure of the low-energy states
of the quantum AKLT Hamiltonian. To wit, we argue that
\beq\label{eq:Psidec}
|\Psi\rangle = \sum_{\alpha = 1}^{\cal N} \sqrt{w_\alpha} \left(\int dg\, g\right)| \Psi_\alpha \rangle
\eeq
up to exponentially small corrections in the thermodynamic limit, where the $|\Psi_\alpha \rangle$ can be interpreted
as a collection of symmetry breaking degenerate quantum ground states whose correlations correspond to the classical pure states $P_\alpha$.

We argue this in three parts. First, we assume the existence of a collection of quantum states $| \Psi_\alpha \rangle$ such that
\begin{equation}
	\label{eq:quant_pure_state}
	|\langle \{ \hat n_i \} | \Psi_\alpha \rangle|^2 = P_\alpha(\{\hat n_i \})
\end{equation}
and show that the quantum state Eq.~\eqref{eq:Psidec} reproduces the observables of the classical decomposition of Eq.~\eqref{eq:PSD}. Second, we will show that up to exponentially small corrections each of the $|\Psi_\alpha \rangle$ are themselves orthogonal ground states. Finally, we address the issue of the existence of such states given the classical distributions $P_\alpha$. 

\newcommand{\OO}{\hat{\mathcal{O}}}

The first part follows the argument of Ref.~\onlinecite{BCZ08} but we  rephrase it in terms of density matrices.
Consider the density matrix of the proposed state Eq.~\eqref{eq:Psidec}:
\beq
	\label{eq:quant_dens_mat}
	\rho = | \Psi \rangle \langle \Psi | = \sum_{\alpha,\beta}\sqrt{w_\alpha w_\beta} \int dg'\,\int dg\, g' |\Psi_\alpha \rangle\langle \Psi_\beta | g^\dagger
\eeq
Given any local operator $\OO$ depending only on spins\footnote{Such observables are diagonal in the coherent state basis and therefore their correlations follow from the classical measure.}, its expectation value in state $| \Psi \rangle$ is given by
\begin{eqnarray}
	\langle \OO \rangle & = & \mbox{Tr}\,\rho\OO \nonumber \\
	& = & \sum_{\alpha,\beta}\sqrt{w_\alpha w_\beta} \int dg'\,\int dg\, \mbox{Tr} \left[ g' |\Psi_\alpha \rangle\langle \Psi_\beta | g^\dagger \OO \right] \nonumber \\
	& = & \sum_{\alpha,\beta}\sqrt{w_\alpha w_\beta} \int dg'\,\int dg\, \langle \Psi_\beta | g^\dagger \OO g' | \Psi_\alpha \rangle
\end{eqnarray}
We now argue that the interference term is negligible. That is,
\beq
\langle \Psi_\beta | g^\dagger \OO g' | \Psi_\alpha \rangle = \delta_{\alpha\beta} \delta_{g\,g'} \langle \Psi_\alpha | g^\dagger \OO g| \Psi_\alpha \rangle
\eeq
up to exponentially small corrections in the thermodynamic limit. This follows from the observation that $|\Psi_\alpha\rangle$ and $|\Psi_\beta\rangle$ have macroscopically distinct magnetization patterns that completely break the $\textsf{O}(3)$ symmetry. In particular, the classical configurations $\{ \hat n_i \}$ on which the wavefunction $| \Psi_\alpha \rangle$ is concentrated have extremely small weight in
any other wavefunction $\beta \neq \alpha$ -- this remains true even with arbitrary global rotations allowed between them. If $\alpha = \beta$ but $g$ and $g'$ differ, then the configurations with weight are again macroscopically distinct by virtue of the net global rotation $g^{-1}g'$ between them. The fact that the observables are local and have bounded matrix elements does not modify these assertions. Since $\mathcal{N}$ is finite for our antiferromagnetic model, the finite sum over the exponentially small corrections does not modify the result:
\beq
\label{eq:quant_exp}
\langle \OO \rangle = \sum_{\alpha}w_\alpha \int dg\, \langle \Psi_\alpha | g^\dagger \OO g | \Psi_\alpha \rangle
\eeq
Inserting a complete set of coherent states and using Eq.~\eqref{eq:quant_pure_state} reproduces the classical distribution Eq.~\eqref{eq:PSD}. The probability of finding the quantum system in a state $\alpha$ (with respect to the state $| \Psi \rangle$) is the same as that of the classical problem:
both are given by $w_\alpha$.

Furthermore, the $| \Psi_\alpha \rangle$ must have exponentially small energy with respect to the quantum Hamiltonian. Since $\langle H \rangle = 0$ in the AKLT state, using equation \eqref{eq:quant_exp} and the rotational invariance of $H$, we have that
\beq
0 = \langle H \rangle = \sum_{\alpha}w_\alpha \langle \Psi_\alpha | H | \Psi_\alpha \rangle
\eeq
up to exponentially small corrections. Since each of the terms in the sum is nonnegative, it follows that
\beq
\langle \Psi_\alpha | H | \Psi_\alpha \rangle \lesssim \mathcal{O}(e^{-N})
\eeq
Thus, the $|\Psi_\alpha\rangle$ are a collection of nearly orthogonal, nearly zero energy states, each of which generates a further continuous collection of such degenerate states under the action of $\textsf{O}(3)$.

Finaly, we turn to the slightly thorny question of whether states satisfying Eq.~\eqref{eq:quant_pure_state} exist. The problem is that the coherent state basis is overcomplete for any fixed spin size $S= z M/2$ and we cannot necessarily find quantum states which have given expansions in this basis. That is, {\it a priori} we cannot simply set $\langle \{ \hat n_i \} | \Psi_\alpha \rangle = \sqrt{P_\alpha(\{ \hat n_i \})}$ and know we have a well-defined quantum state for spins of size $S$. In the large spin limit, there is no problem as the coherent states become a complete, rather than overcomplete, basis. This coincides with the zero temperature limit of the classical companion model and thus the pure states $| \Psi_\alpha \rangle$
may be identified with the (many degenerate) minima of the energy function (\ref{eq:aklt-ham}). At finite $M$, we cannot find such finely localized states in the coherent state representation -- the most localized state has solid angular scale $\sim 1/M$ -- but the finite temperature fluctuations around the classical minima will also smear the $P_\alpha$ at a similar scale. 

On the other hand, we already used above that the states have disjoint support, up to exponentially small corrections in $N$.
For a given classical configuration $\{ \hat n_i \}$, only one state contributes to $\langle \{ \hat n_i \} | \Psi \rangle$ significantly. Thus, in a given region of classical configuration space,
$\langle \{ \hat n_i \} | \Psi \rangle$ coincides with one of the $\langle \{ \hat n_i \} | \Psi_\alpha \rangle$, and conversely, each of the $\langle \{ \hat n_i \} | \Psi_\alpha \rangle$ can be seen
as the restriction of the full $\langle \{ \hat n_i \} | \Psi \rangle$ to  that region. Since the $1/M$ smoothing is local in configuration space, it is safe to assume that
the states $\langle \{ \hat n_i \} | \Psi_\alpha \rangle$ are as smooth as the original $\langle \{ \hat n_i \} | \Psi \rangle$, therefore $| \Psi_\alpha \rangle$ can be defined without ambiguity.
Based on the above arguments, we think it likely that at least an approximate pure state decomposition of the form proposed above can be found even at finite $M$.

In summary, we obtain the following picture for the low energy spectrum in the spin glass phase: the non-clustering paramagnetic AKLT ground state $| \Psi \rangle$ can be decomposed in
a superposition of several almost degenerate states, whose energies are of order $\exp(-N)$. These states enjoy the clustering property (vanishing of connected correlations) and are
characterized by amorphous order (the local magnetizations are different in each state). It would be nice to check this scenario explicitly by means of exact diagonalization.

It would be very interesting to obtain more detailed information on the spectrum of such spin glass Hamiltonians. For instance, a natural question is whether there is an energy gap between the degenerate
low-lying spin glass states and the excited states. Indeed, we expect Goldstone (or Halperin-Saslow) modes\cite{Halperin:1977vn} associated with twisting of the amorphously magnetized states $\ket{\Psi_\alpha}$. While it is difficult to explicitly construct a coarse-graining procedure to produce an effective theory of such modes on a tree-like graph, one usually expects that such a theory applies in sufficiently high dimensions\cite{Gurarie:2003p8714}. Insofar as the effective theory is an elastic hydrodynamics living on a tree-like graph, the corresponding modes should remain gapped\cite{Laumann:2009p8702}. This suggests that the low energy spectrum is indeed gapped in any given pure state sector.


\section{Concluding Remarks} 
\label{sec:conclusions}

In this paper, we extend the study of AKLT models to locally tree-like graphs of fixed connectivity by exploiting the quantum-classical mapping of the associated wavefunctions. On the infinite Cayley tree, we recover the results obtained in  Ref.~\onlinecite{Fannes:1992p5750} . We find that the Bethe lattice possesses the peculiar property that it is possible to choose parameters (for $z=3,4$) so that the corresponding AKLT state is \emph{critical}. A variational calculation of the gap is unable to produce gaplessness, which is consistent with the arguments of Ref.~\onlinecite{Laumann:2009p8702} that this is a general feature of locally tree-like graphs: essentially, one cannot deform a uniform excitation into long-wavelength rotations of the order parameter, without jumping a gap in the Laplacian spectrum.

Turning to regular random graphs, we find that the companion classical model is unstable to spin glass ordering within a cavity analysis. This is a general feature of classical antiferromagnetic models on such graphs, but has  striking consequences given that the pecularities of such mean-field-like glasses should directly transfer to the \emph{quantum} ground state of the AKLT model. This provides an alternative route to the study of quantum glassy order in tree-like models (see for example Ref.~\onlinecite{Laumann:2008kx,Kopec:2006p7443,Kopec:1997p341,TB08,CTZ09,XORSAT}). 
We argue that there are now many (nearly) degenerate quantum ground states with macroscopically distinct magnetization patterns, but that there remains a gap to Halperin-Saslow waves for geometric reasons analogous to the simpler case of the antiferromagnet. 

There are several avenues for future research. One obvious direction is to study the classical vector spin glass and the corresponding classical measure.  In a different vein, we observe that the AKLT construction applies at a very special point in the space of quantum Hamiltonians. To what extent do the features of the quantum AKLT glass extend to regions proximate to this exactly solvable point? Ideally, the AKLT glass would capture the essential features of a broader range of quantum spin glasses, playing a role reminiscent of  that played by the $S=1$ AKLT chain in relation to the Haldane phase.


\section*{Acknowledgements} 

CRL and SAP thank the hospitality of the \'Ecole de Physique des Houches where this work was initiated while running down locally tree-covered mountain paths. We would also like to thank Michael Aizenmann for several stimulating discussions, and for directing us to the literature on the spectral properties of random graphs. 
SAP and SLS would like to thank Dan Arovas for introducing them to the challenges of studying AKLT models in
general dimensions and for collaboration on other work in this area.
FZ wishes to thank the Princeton Center for Theoretical Science for hospitality during most of this work.


\appendix

\section{Transfer Matrix for the AKLT Model} 
\label{appsec:Tmatrix}
The transfer matrix may be thought of as a map between functions
defined on the sphere:
\begin{equation}
(Tf) (\nhat) = \int \frac{d\nhat'}{4\pi} T(\nhat, \nhat') f(\nhat')
\end{equation}
where $\nhat, \nhat' \in S^{N-1}$. In our case, as in most cases of
interest, the transfer matrix is a
rotational scalar and we can work in the angular momentum basis. The
eigenvalue must depend only on the $L^2$
eigenvalue $l$ and not on the $L_z$ eigenvalue $m$. It therefore
suffices to solve the problem in the case $m=0$.

We wish to solve the eigenvalue equation:
\begin{equation}\label{eq:basiceigenvalue}
\lambda_l  f_l(\nhat) = \int_{S^{2}} \frac{D\nhat'}{4\pi}\, T(\nhat, \nhat') f_l(\nhat')
\end{equation}
Since the kernel $T$ depends only on
$\nhat\cdot\nhat' =\cos\theta$, we work in polar coordinates with the $z$-axis along $\nhat$ and substitute $x=\cos\theta$ to obtain
\begin{equation}
\lambda_l f_l(1) =  \frac{1}{2}\int_{-1}^{1} dx \, T(x) f_l(x)
\end{equation}
A natural guess for the eigenfunctions is that they are Legendre
polynomials. The transfer matrix in our case is $T(x) =
(\frac{1+x}{2})^{\beta}$.Using standard identities,
\begin{eqnarray}\label{eq:finaleval}
\lambda_{l, \text{AKLT}} &=& 
 \frac{\left[\Gamma(\beta+1)\right]^2}{\Gamma(\beta -l +1) \Gamma(\beta+l+2)}
\end{eqnarray}
A similar discussion for the Heisenberg model for arbitrary $N$ and the case of $SU(N)$ and $Sp(N)$ groups, may be found in
Ref.~\onlinecite{Fendley:2002p5640}.


\section{Stability against spin glass ordering on a regular random graph} 
\label{sec:stabSG}

We follow closely the derivation of Ref.~\onlinecite{KZ07}, Appendix A. In the thermodynamic limit, the spin glass susceptibility
\beq
\chi_{SG} = \frac1N \sum_{ij} [ \langle \hat n_i \cdot \hat n_j \rangle ]^2
\eeq
can be rewritten, 
by taking the average over the random graphs and using translational invariance, as
\beq
\chi_{SG} = \sum_{d=0}^\infty {\cal N}_d [ \langle \hat n_0 \cdot \hat n_d \rangle ]^2
\eeq
where ${\cal N}_d$ is the number of sites at distance $d$ from a reference site.
The sum is convergent as long as
\beq
\lim_{d\to\infty} ({\cal N}_d)^{1/d} [ \langle \hat n_0 \cdot \hat n_d \rangle ]^{2/d} \leq 1
\eeq
Note that $({\cal N}_d)^{1/d} \to z-1$ for large $d$.
In the paramagnetic phase, $\langle \hat n_0 \cdot \hat n_d \rangle$ is given by the response of $\langle \hat n_0 \rangle$
(the root) to a small magnetic field coupled to $\hat n_d$, a leaf at distance $d$, of a tree whose other nodes are in the paramagnetic state $|00\rangle$.
Hence we get (repeated indices summed):
\begin{equation}	
\langle \hat n_0 \cdot \hat n_d \rangle \propto \frac{d \langle \hat n_0^i \rangle}{d h_d^i}
= \int d \hat n_0\, \hat n_0^i \frac{d \psi^0(\hat n_0)}{d h_d^i}
\end{equation}
Clearly the term that gives the exponential dependence on $d$ is the variation of $\psi^0(\hat n_0)$ with respect to $h_d$.
Using the recursion relation (\ref{eq:cont-cavity-iter}) we can rewrite it as
\begin{eqnarray*}
	\frac{d \psi^0(\hat n_0)}{d h_d} &=& \int d \hat n_1 \cdots d \hat n_d  \frac{d \psi^0(\hat n_0)}{d \psi^1(\hat n_1)}
	\frac{d \psi^1(\hat n_1)}{d \psi^2(\hat n_2)}
	\cdots \frac{d \psi^{d}(\hat n_d)}{d h_d}	
\end{eqnarray*}
and the exponential dependence is related to the eigenvalue of the transfer matrix
$\frac{d \psi^d(\hat n_d)}{d \psi^{d+1}(\hat n_{d+1})} = T(\hat n_d, \hat n_{d+1})$.
These can be obtained by repeating the analysis of section~\ref{sec:cont-spin-model}. Indeed if we use the same ket notation and
write the variation $\delta \ket{d+1} = \epsilon\sum_{l\neq0,
  m}^{\infty} c_{lm}\ket{l\,m}$ we can use the results of section~\ref{sec:cont-spin-model} to obtain
\beq
\delta \ket{d} = \epsilon \sum_{l\neq 0, m} \frac{\lambda_l}{\lambda_0}c_{lm}\ket{l\,m} + \mathcal{O}(\epsilon^2)
\eeq
The absence of the factor $(z-1)$ with respect to Eq.~(\ref{eq:cont-iter-state}) is due to the fact that
here we are only varying one of the neighbors of a given spin, the neighbor on the path linking the root to the given
leaf at distance $d$. The relevant eigenvalues are therefore $\lambda_l/\lambda_0$ and we obtain the condition
\beq
(z-1) \max \{ (\lambda_l/\lambda_0)^2 \} \leq 1
\eeq
for the convergence of $\chi_{SG}$.


\bibliography{aklt-papers}

\end{document}